\documentclass[12pt]{article}

\usepackage{epsfig} 
\usepackage{amsfonts}
\usepackage{graphicx}

 \usepackage{amssymb}

\date{\today}

\newcommand{\insertplot}[5]{\begin{figure}
 \hfill\hbox to 0.05in{\vbox to #5in{\vfill
 \inputplot{#1}{#4}{#5}}\hfill}
 \hfill\vspace{-.1in}
 \caption{#2}\label{#3}
 \end{figure}}
 \newcommand{\inputplot}[3]{
 \special{ps: plotfile #1}
}
\newcounter{fig}

\renewcommand{\t}{\theta}

\newcommand{\ee}{\end{equation}}
\newcommand{\eea}{\end{eqnarray}}
\newcommand{\be}{\begin{equation}}
\newcommand{\bea}{\begin{eqnarray}}

\begin{document}

\title{ 
Spinning gauged boson stars in anti-de Sitter spacetime
} 
  
\author{
{\large Olga Kichakova}, 
{\large Jutta Kunz} 
and {\large Eugen Radu}  \\ 
 {\small  Institut f\"ur Physik, Universit\"at Oldenburg, Postfach 2503
D-26111 Oldenburg, Germany} }

\maketitle

\begin{abstract} 
We study axially symmetric  solutions of the 
 Einstein-Maxwell-Klein-Gordon equations 
describing spinning gauged boson stars in a 3+1 dimensional
asymptotically AdS spacetime. 
These smooth horizonless solutions possess an electric charge and a magnetic dipole moment, their
 angular momentum being proportional to the electric charge.
 A special class of solutions with a self-interacting scalar field,
  corresponding to static axially symmetric solitons with
a nonzero magnetic dipole moment, is also investigated. 
\end{abstract}

\section{Introduction}

 Recently there has been a lot of interest in  
 solutions of the general relativity with a negative cosmological constant $\Lambda$
coupled to a Maxwell field and a charged scalar with mass $M$ and gauge coupling constant $q$.
  Working in four spacetime dimensions, 
this system is usually described by the action 
\begin{eqnarray} 
\label{action}
I=\int_{\cal M} d^{4}x\sqrt{-g}\left [\frac{1}{16\pi G} (R-2\Lambda)
- \frac{1}{4} F_{\mu\nu} F^{\mu\nu} -
\frac{1}{2}g^{\mu \nu} \left( D_\mu\psi^* D_\nu \psi+D_\nu\psi^* D_\mu \psi \right) - U(|\psi|) \right] ,~{~}
\end{eqnarray}
where $G$ is the gravitational constant, $R$ is the Ricci scalar associated with the
spacetime metric $g_{\mu\nu}$,
 $F_{\mu\nu} =\partial_\mu A_\nu - \partial_\nu A_\mu$ is the U(1) field strength tensor and
$D_\mu\psi=\partial_\mu \psi + iq A_\mu \psi$ is the covariant derivative.
  $ U(|\psi|)$ denotes the potential of the scalar field $\psi$,
whose mass is defined by  $M^2=\frac{1}{2}\frac{\partial^2 U}{\partial |\psi|^2}\big|_{\psi=0}$.
 
Although (\ref{action})
does not correspond to a consistent truncation of a
more fundamental theory 
(unless  $\psi$ is vanishing), it can be viewed however, 
as a simple toy model for the charged scalar dynamics of systems
that appear in concrete examples of the AdS/CFT correspondence.
 The interest in this system 
(sometimes called the gravitating Abelian Higgs model) 
has been boosted  due to Gubser's observation \cite{Gubser:2008px} 
that
the Einstein-Maxwell black holes
can become unstable to forming scalar hair at low temperatures. 
This instability results in new branches of solutions with scalar hair,
which are thermodynamically 
favoured over the Reissner-Nordstr\"om-AdS black holes.
The charged
black branes which are asymptotically AdS in a Poincar\'e coordinate patch
are of main interest,
their physics being recently exploited to obtain a dual gravitational 
description of important phenomena in condensed matter
physics (in particular superconductivity and phase transitions)  in a three dimensional flat spacetime $R_t\times R^2$. 
 A discussion of these aspects can be found $e.g.$ in \cite{Horowitz:2010gk},
 together with a large set of references.

However, another natural arena to study solutions of 
the model (\ref{action}) is to 
consider instead a globally AdS background, with a line element
$
ds^2 =-(1+\frac{r^2}{\ell^2}) dt^2 
+\frac{dr^2}{1+\frac{r^2}{\ell^2}} + r^2 (d\t^2+\sin^2 \t  d \varphi^2), 
$
(where $t$ is the time coordinate and  $r$, $\theta$, $\varphi$ are
the spherical coordinates with the usual range and $\Lambda=-3/\ell^2$),
the conformal boundary being a static Einstein universe  $R_t\times S^2$.
In this case, the simplest nonvacuum solutions of the model (\ref{action}) 
are the boson stars with a vanishing gauge field, $F_{\mu \nu}=0$. 
These are smooth horizonless configurations representing
gravitational bound states of a complex scalar field with a harmonic 
 time dependence, which provide us 
with the simplest model of a relativistic star\footnote{One should remark that in contrast to 
 ordinary stars or neutron stars, generically
the boson stars do not display a
sharp edge. In this case, the matter is not confined in a finite region of space and
the boson stars possess only an {\it effective} radius \cite{Jetzer:1991jr} 
(see however, the compact boson stars in \cite{Kleihaus:2009kr}).}.
Such solutions have been extensively studied for $\Lambda=0$,
$i.e.$ an asymptotically flat spacetime background, starting with the early work
of Kaup \cite{Kaup:1968zz} and Ruffini and Bonazzola 
\cite{Ruffini:1969qy}.
They have found interesting physical applications, being proposed as
candidates for dark matter halos and as
 dark alternatives to astrophysical black hole candidates; also,  they may help explain galaxy rotation curves 
-- see the  review work \cite{Jetzer:1991jr}.

The boson starts with AdS asymptotics  
are believed to play  an important role in holographic gauge theories
through the AdS/CFT correspondence\footnote{However, note that the boson starts do not have natural counterparts 
in a Poincar\'e coordinate patch.  
}.
While the early studies restricted to spherically symmetric configurations
\cite{Astefanesei:2003qy},
\cite{Prikas:2004yw} (see also \cite{Buchel:2013uba,Hartmann:2013kna})
recently there has been
some progress on including the effects of rotation.
Spinning AdS boson stars in $d=4$ spacetime dimensions have been discussed in \cite{Radu:2012yx}.
These stationary localized configurations possess a finite mass and angular momentum, 
their angular momentum being quantized,
$J=n Q$ (with $n$ an integer and $Q$ the Noether charge), 
the energy density exhibiting a toroidal distribution.
A particular set of higher dimensional\footnote{For completeness, we mention that spinning boson stars 
with an AdS$_3$ background have been studied in \cite{Astefanesei:2003rw}.
However, these solutions have rather special properties.} spinning boson stars possessing equal angular momenta has been considered in 
\cite{Dias:2011at}, 
\cite{Stotyn:2011ns},
for $d=2k+1\geq 5$ and a special multiplet scalar fields ansatz 
\cite{Hartmann:2010pm}.

An interesting question to address in this context is the issue of rotating
horizonless solutions within the full model (\ref{action}),
and, in particular, how an electric charge
would affect their properties.
For $\Lambda=0$, this  problem has been addressed in Ref. \cite{Brihaye:2009dx},
which gave numerical evidence for the existence of spinning gauged boson stars in 
a Minkowski spacetime background.
Similar to a Kerr-Newman black hole, these solutions possess a
nonzero electric charge and a magnetic dipole moment.
However, their pattern is rather similar to that of the ungauged boson stars discussed in 
\cite{Schunck:1996wa},
\cite{Yoshida:1997qf},
\cite{Kleihaus:2005me};
in particular, one finds again a
limited range for the allowed scalar field frequencies.

In this work we show that the 
 spinning gauged boson stars in \cite{Brihaye:2009dx}
can be generalized for an AdS background.
In some sense, these globally regular configurations can be regarded as regularized 
Kerr-Newman-AdS solutions, 
the event horizon and the singularity disappearing due to the supplementary
interaction with a complex scalar field.
These gauged boson stars can also be viewed as
axially symmetric generalizations of the spherically 
symmetric gauged solutions in 
\cite{Prikas:2004yw},
\cite{Gentle:2011kv},
\cite{Hu:2012dx},
\cite{Nogueira:2013if}.
Similar to that case,
our results show that they exist also up to a maximal value of the 
gauge coupling constant. Their angular momentum is 
quantized, being proportional to the electric charge.

This paper is organized as follows.
In the next Section
we formulate a numerical approach of the problem based on a specific ansatz. 
The numerical results are given in Section 3, where we
exhibit the physical properties of the gauged spinning boson star solutions.
We conclude in Section 4
with some further remarks.

\section{The problem}

\subsection{The equations and the ansatz  }

The configurations investigated in this work are solutions of the coupled
Einstein-Maxwell-scalar field equations
 \begin{eqnarray}
\label{field-eqs}
 &&E_{\mu\nu}=R_{\mu\nu}-\frac{1}{2}g_{\mu\nu}R+\Lambda g_{\mu\nu} -8 \pi G T_{\mu\nu}=0\ , 
\\
\label{scalar-eqs} 
&&
D_{\mu}D^{\mu}\psi=\frac{\partial U}{\partial\left|\psi\right|^2} \psi,
~~\nabla_{\mu}F^{\mu\nu}=
iq \big [ (D^{\nu}\psi^*) \psi-\psi^*(D^\nu \psi) \big ]~
\equiv q j^\nu ~,
\end{eqnarray}  
where $T_{\mu\nu}$ is the
 stress-energy tensor 
\begin{eqnarray}
\label{tmunu} 
T_{\mu \nu}  
&=&
F_{\mu \alpha}F_{\nu \beta}g^{\alpha \beta}-\frac{1}{4}g_{\mu\nu}F_{\alpha\beta}F^{\alpha\beta}
\\
\nonumber
&&
+\left(
 D_\mu\psi^*D_\nu\psi
+ D_\nu\psi^*D_\mu\psi
\right )
-g_{\mu\nu} \left[ \frac{1}{2} g^{\alpha\beta} 
\left( D_\alpha\psi^* D_\beta \psi+D_\beta\psi^* D_\alpha \psi \right)+U(\left|\psi\right|)\right]
 \ .
\end{eqnarray} 
This model is invariant under the local U(1) gauge transformation 
\begin{eqnarray}
\label{gauge-transf}
\psi \to \psi e^{-i q \alpha},~~A_\mu\to A_\mu +\partial_\mu \alpha,
\end{eqnarray}
with $\alpha$ a real function.


We are interested in stationary axially symmetric configurations, 
with a spacetime geometry admiting two Killing vectors 
$\partial_t$
and 
$\partial_{\varphi}$,
in a system of adapted coordinates. 
Then the most general line element is written as
$
ds^2 = G_{tt}(x) dt^2+  2G_{t\varphi}(x) dt d\varphi+ G_{\varphi\varphi}(x) d\varphi^2
+h_{ij}(x)dx^i dx^j, 
$
with $x^i=(r,\theta)$. 
In the numerics, it is convenient to choose a metric gauge with 
$(1+\frac{r^2}{\ell^2})h_{rr}=h_{\theta \theta}/r^2 $ and $h_{r\theta}=0$.
This leads to a metric ansatz with four unknown functions,
a convenient form being 
\begin{eqnarray}
\label{ansatz-metric}
&
ds^2 =- F_0(r,\theta) (1+\frac{r^2}{\ell^2}) dt^2 
+ F_1(r,\theta) \left( \frac{dr^2}{1+\frac{r^2}{\ell^2}} + r^2 \, d\t^2 \right) 
  + F_2(r,\theta) r^2 \sin^2 \t  \left( d \varphi
- \frac{W(r,\theta)}{r}   dt \right)^2 ,{~~~~}
\end{eqnarray}
with  
$ (F_i, W)$ (where $i=0,1,2)$
being smooth
in $r,\theta$.

For the scalar field,  we adopt the stationary ansatz 
\cite{Schunck:1996wa},
\cite{Yoshida:1997qf},
\cite{Kleihaus:2005me}:
\begin{eqnarray}
\label{ansatz-scalar}
\psi (t,r,\t, \varphi)= \phi(r, \t)
 e^{ i( n \varphi-\omega  t )}   , 
\end{eqnarray}
where $\phi(r, \theta)$ is a real function,
and $\omega $ and $n$ are real constants.
Single-valuedness of the scalar field requires
$\psi(\varphi)=\psi(2\pi + \varphi)$;
thus the constant $n$ must be an integer,
$i.e.$, $n \, = \, 0, \, \pm 1, \, \pm 2, \, \dots$~.
In what follows, we shall take $n\geq 0$ and $\omega \geq 0$,
without any loss of generality.  

A consistent ansatz for the U(1) gauge field
reads 
\begin{eqnarray}
\label{ansatz-U1}
{\cal A}= A_\mu dx^\mu=A_t(r,\theta) dt+ A_{\varphi}(r,\theta)\sin \theta (d\varphi-\frac{W}{r} dt).
\end{eqnarray}

Substituting 
(\ref{ansatz-metric}), 
(\ref{ansatz-scalar}), 
(\ref{ansatz-U1}) 
in the field equations (\ref{field-eqs}), (\ref{scalar-eqs})
results\footnote{Note that the Einstein equations $E_\theta^r =0,~E_r^r-E_\theta^\theta  =0$ are not automatically satisfied,
yielding two constraints. However, following  \cite{Wiseman:2002zc}, one can show that
these constraints are satisfied as a consequence of the identities $E_{\mu; \nu}^{\nu}=0$  plus the
set of chosen boundary conditions.
In practice, the constraint equations $E_\theta^r $ and $E_r^r-E_\theta^\theta$ are used 
to monitor the numerical accuracy of the solutions.} 
in a set of seven coupled non-linear PDEs of the form
$\nabla^2 {\cal F}_a={\cal J}_a$  
where ${\cal F}_a=(F_0,F_1,F_2,W; \phi; A_{\varphi},A_t)$, 
${\cal J}_a$ are `source' terms depending on the functions ${\cal F}_a$ and their first derivatives, while
$\nabla^2$ is the Laplace operator associated with the auxiliary space
$d\sigma^2=dr^2/(1+\frac{r^2}{\ell^2})+r^2 d\theta^2$.

Solutions of this model with a vanishing gauge field $A_t=A_{\varphi}=0$
have been discussed in \cite{Radu:2012yx},
generalizing for the AdS spacetime the asymptotically flat rotating boson stars in 
\cite{Schunck:1996wa},
\cite{Yoshida:1997qf},
\cite{Kleihaus:2005me}.
Note that, in contrast to that case,  the $(t, \varphi)$-dependence of the scalar field $\psi$
can now be gauged away by applying the local U(1) symmetry
(\ref{gauge-transf})
with $\alpha =  (n\varphi -\omega t)/q$. 
However, this would also change the gauge field as $A_t\to A_t-\omega/q$,
$A_\varphi \to A_\varphi+n/q$, so that it would become singular in the $q\to 0 $ limit. 
Therefore, in order
to be able to consider this limit, we prefer to keep the $(t,\varphi)$-dependence in 
the scalar field ansatz and to fix
the corresponding gauge freedom by setting $A_t = A_\varphi = 0$ at infinity.
 
 We note also that the spherically symmetric limit is found for $n=0$, in which case the
functions $F_0,F_1,F_2$ and $\phi$, $A_t$
depend  only on $r$, with $F_1=F_2$ and $W=A_\varphi=0$.

\subsection{The asymptotics and boundary conditions}
We are interested in  horizonless, particle-like solutions 
of the equations (\ref{field-eqs}), (\ref{scalar-eqs})
within the ansatz (\ref{ansatz-metric}), (\ref{ansatz-scalar}), (\ref{ansatz-U1}),
  approaching at infinity
the globally AdS background. 
Since this problems does not seem to possess closed form solutions,
the field equations are solved numerically with suitable boundary conditions.
These conditions result  from a study of an approximate form of the solutions
on the boundaries of the domain of integration,
compatible with regularity and AdS asymptotics requirements.  

For small values of $r$, the solutions  possess a power series  on the form
$F_k=F_{k0}+O(r^2)$ (with  $k=0,1,2$ and 
$F_{10}=F_{20}$), $W=O(r^2)$, 
$\phi=O(r^{2n})$, 
$A_{\varphi}=O(r^2)$ and 
$A_t=V+O(r^2)$.
This leads to the following
 boundary conditions at the origin:
\begin{eqnarray}
\label{bc0} 
\partial_r F_i|_{r=0}= 
W|_{r=0}=0,~~
\phi| _{r =0}=0,~~\partial_r A_t|_{r=0}=A_\varphi|_{r=0}=0.
\end{eqnarray}

At infinity, the AdS background is approached, while the scalar field\footnote{Without any loss of generality,
we suppose $U(\phi)\to 0$ as $\phi\to 0$.} $\phi$ 
and the gauge potential $A_\mu$ vanish. 
Restricting to the case $M^2\geq 0$,  
the matter fields decay asymptotically as 
 \begin{eqnarray}
 \label{asym-matter-fields}
\phi\sim \frac{c_1(\theta)}{r^\Delta}+\dots~,~~
A_t\sim 
\frac{Q_e}{r}+\dots~,~~~A_{\varphi}\sim \frac{\mu \sin \theta}{r}+\dots,
 \end{eqnarray}
where
$Q_e$, $\mu$ and $c_1(\theta)$ result  from numerics,
while 
$
\Delta=\frac{3}{2}\left( 1+\sqrt{1+\frac{4}{9} M^2 \ell^2} \right).
$
Then the Einstein equations imply the following form of the metric functions
as $r\to \infty$
 \begin{eqnarray}
\label{asym1}
&F_0=1+ \frac{f_{03}(\theta)}{r^3}+ \dots,
~F_1=1+ \frac{f_{13}(\theta)}{r^3}+\dots, 
~F_2=1+ \frac{f_{23}(\theta)}{r^3}+\dots,
~W=\frac{w_2(\theta)}{r^2}+\dots,~~~{~~~}
\end{eqnarray}
in terms of two functions $f_{13}(\theta)$ and $w_2(\theta)$ which result from the numerics, 
with 
%
$
f_{03}(\theta)=-3 f_{13}(\theta)-\frac{4}{3}\tan\theta f_{13}'(\theta),
$ 
and
$
f_{23}(\theta)=  f_{13}(\theta)+\frac{4}{3}\tan\theta f_{13}'(\theta).
$
%
Then the corresponding boundary conditions employed in the numerics are
\begin{eqnarray}
\label{bcinf} 
F_i|_{r \rightarrow \infty} =1,~~
W|_{r \rightarrow \infty} =0, ~~
\phi| _{r \rightarrow \infty}=A_t|_{r \rightarrow \infty}=A_\varphi|_{r \rightarrow \infty}=0 .
\end{eqnarray}
For $\t=0$,  
the study of an approximate 
solution of the equations (\ref{field-eqs}), (\ref{scalar-eqs}) implies the behaviour
$F_k=\bar F_{k0}(r)+O(\theta^2)$,
 $W=w_0(r)+O(\theta^2)$, $\phi=O(\theta^{2n})$, $A_{\varphi}=O(\theta)$ and $A_t=A_t^0(r)+O(\theta^2)$.
A similar expansion holds also for $\theta = \pi$, with $\theta\to \pi-\theta$.
Thus we require the boundary conditions
\begin{eqnarray}
\label{bct0} 
\partial_{\t} F_i|_{\t=0,\pi}= 
\partial_{\t} W |_{\t=0,\pi}=0,~~
\phi |_{\t=0,\pi}= \partial_{\t}A_t|_{\t=0,\pi}=A_\varphi |_{\t=0,\pi}=0.
\end{eqnarray}
 The absence of conical singularities
  imposes on the symmetry axis the supplementary condition 
$F_1|_{\theta=0,\pi}=F_2|_{\theta=0,\pi},$
which is  used to verify the accuracy of the solutions.

Also, all solutions in this work 
are invariant\footnote{However, gauged boson stars with odd parity with respect to 
a reflection in the equatorial plane should also exist.
For $\Lambda=0$ and a vanishing gauge field, 
such configurations have been studied in \cite{Kleihaus:2007vk}.} 
under the parity transformation $\theta \to\pi-\theta $.
We make use of this symmetry to integrate the equations for $0\leq \theta\leq \pi/2$ only, the
following boundary conditions being imposed in the equatorial plane
\begin{eqnarray}
\label{bctpi2} 
\partial_{\t} F_i|_{\t=\pi/2}= ~
\partial_{\t} W |_{\t=\pi/2}=0 \ ,~~
\partial_{\t} \phi |_{\t=\pi/2}=\partial_{\t}A_t|_{\t=\pi/2}=\partial_{\t}A_\varphi|_{\t=\pi/2}=0 \ .
 \end{eqnarray}

\subsection{The global charges}
The charged boson star solutions possess two global charges associated with
the asymptotic Killing vectors  
$\partial_t$
and 
$\partial_{\varphi}$ of the metric, which are the mass-energy $E$ and 
angular momentum $J$.
To compute these quantities, we employ first the 
quasilocal formalism in 
 \cite{Balasubramanian:1999re}. 
 In this approach, the action (\ref{action}) is supplemented with the Gibbons-Hawking surface term  
 \cite{Gibbons:1976ue}
 and a  boundary 
 counterterm $I_{\rm ct}=-\frac{1}{8 \pi G} \int_{\partial {\cal M}}d^{3}x\sqrt{-h}
 (
\frac{2}{ \ell}+\frac{ \ell}{2}\rm{R}
)$  
(where $\rm{ R}$ is the Ricci scalar for the boundary metric $h$).
Then the variation of the total action with respect to $h_{ab}$
results  \cite{Balasubramanian:1999re} in the   boundary stress tensor  
$
{\rm T}_{ab} 
=\frac{1}{8\pi G}(K_{ab}-Kh_{ab}-\frac{2}{\ell}h_{ab}+\ell E_{ab}) 
$
(where $E_{ab}$ is the Einstein tensor of the boundary metric, $K_{ab}$
is the extrinsic curvature
tensor of the boundary and $K=h_{ab}K^{ab}$).
In this approach, 
the mass-energy and angular momentum  are computed from ${\rm T}_{ab}$,
being regarded as conserved charges associated with 
the Killing vectors  $\partial/\partial t$ and $\partial/\partial \varphi$ of the boundary metric.
For the metric ansatz (\ref{ansatz-metric}) (with the
asymptotics (\ref{asym1})), a straightforward computation leads to the following expressions
of these quantities
  \begin{eqnarray}
\label{EJ}
E= \frac{1}{8 G \ell^2}\int_{0}^\pi d\theta \sin\theta \bigg(5 f_{13}(\theta)+3 f_{23}(\theta)\bigg),~~
J=-\frac{3}{8 G }\int_{0}^\pi d\theta \sin^3\theta ~w_2(\theta) .
\end{eqnarray}   
A similar result is found when using instead the Ashtekar-Magnon-Das formalism in 
\cite{Ash}.
 Moreover, 
the same expression for the angular momentum is found from the Komar integral,
$J=\frac{1}{8 \pi G }\int R_{\varphi}^t \sqrt{-g} dr  d\theta d\varphi$. 

Apart from $E$ and $J$, the system possesses two global  charges associated with the matter fields.
These are the Noether charge ($i.e.$ the total
particle number)
\begin{eqnarray}
\label{Q1}
Q= \int j^t \sqrt{-g} dr  d\theta d\varphi=
2 \pi \int_0^\infty dr \int_0^\pi d\theta~
2r^2 \sin \theta F_1\sqrt{\frac{F_2}{F_0}}
\frac{ \phi^2}{1+\frac{r^2}{\ell^2}}(\omega -q  A_t-\frac{n W}{r}),
\end{eqnarray} 
and the electric charge $Q_e$ which is read from the asymptotics
 of the electric potential $A_t$ as given in  (\ref{asym-matter-fields}).
However, 
a straightforward computation 
shows that both the Noether charge and the electric charge of the spinning solutions 
are proportional\footnote{This relation is valid also for
spinning solutions with a magnetic dipole
in a generalization of (\ref{action})
with an SU(2) local gauge symmetry of the matter lagrangian.
However, in contrast to (\ref{action}),
a Yang-Mills-Higgs theory possesses also solutions
with a non-zero magnetic flux at infinity.
As discussed in \cite{vanderBij:2002sq}, the total angular momentum of such non-Abelian configurations is zero.  
} 
to the total angular momentum,
\begin{eqnarray}
\label{JQ}
J= n Q=\frac{Q_e n}{q}.
\end{eqnarray}  
 These configurations possess also a magnetic dipole moment $\mu$ which is read from the 
 asymptotics  (\ref{asym-matter-fields}) of the magnetic gauge potential.

The gauged spinning boson stars have no horizon and therefore they are zero entropy objects,
without an intrinsic temperature.
The first law of thermodynamics
reads in this case 
$dE= \omega dQ=\frac{\omega }{n} dJ=\Phi  dQ_e$,
with
$\Phi=\omega/q$ the  electrostatic potential.

 \section{The results}
In the numerics, we set $4\pi G=1$ and express all variables and quantities
in natural units set by the AdS length scale $\ell$  
(for example, we 
scale $r\to r/\ell$, $\omega\to \omega \ell$, $M\to M/\ell$ etc.).

The set of seven coupled non-linear
elliptic partial differential equations
for the functions ${\cal F}_a$
has been solved numerically  
subject to the boundary conditions 
(\ref{bc0}) at $r=0$,
(\ref{bcinf}) at infinity,
(\ref{bct0}) for $\theta=0$,
and respectively
(\ref{bctpi2}) in the equatorial plane.
In practice, we have
compactified $r$ to the coordinate $x\in [0,1]$,
where $dr=\frac{dx}{(1-x)^2}$.
Also,  the field equations have been discretized
using a fourth order finite difference scheme,
taking a uniform grid with $N_x\times N_\theta$ points 
(typically $N_x=250$, $N_\theta=30$).
All numerical calculations have been 
performed by using the programs FIDISOL/CADSOL \cite{schoen}.
This software{\footnote{We have found this method robust and allowing
for an accurate extraction of the boundary quantities.
The Ref. \cite{schoen} 
provides a detailed description of  the numerical method and explicit examples
(see also $e.g.$ the Appendix in \cite{Kleihaus:2011yq} for a discussion within a physical problem).}
 provides also an
error estimate for each unknown function. The typical relative error
for the solutions reported in this work is estimated to be of the order of $10^{-3}$.

In our approach, the input data
are the scalar potential 
$U(|\psi|)$, 
the frequency $\omega$ 
and
the winding number 
$n$ in the ansatz (\ref{ansatz-scalar}) for the scalar field $\psi$,
the gauge coupling constant $q$ and the AdS length scale $\ell$.
All quantities of interest ($e.g.$ the mass-energy $E$ and angular momentum $J$) are extracted 
from the numerical solutions.
Also, for 
  simplicity we restrict our study in this work to the case of a 
  nodeless scalar field\footnote{However,
angular excited boson stars are also likely to exist,
representing excited states of the model.
Their basic properties
case have been discussed in \cite{Brihaye:2009dx}, for an asymptotically flat
spacetime.}.

\subsection{The probe limit: spinning gauged Q-balls in a fixed AdS background}

Before discussing the properties of the gravitating solutions, it is 
useful to consider first the so-called `probe limit'.
In this case, the backreaction  on the spacetime geometry is ignored
and the gauged scalar field solutions are studied in a fixed AdS background, $i.e.$
with $F_1=F_2=F_0=1,~W=0$ in (\ref{ansatz-metric}).
Then the problem reduces to solving three PDEs for the matter functions $\phi,A_\phi$ and $A_t$.
These solutions describe a special class of non-topological solitons--the
U(1)-gauged Q-balls 
(see Ref.  \cite{Radu:2008pp} for a discussion of 
the corresponding problem for $\Lambda=0$).
The scalar field possesses in this case a self-interaction potential, the usual form in the
literature being (see $e.g.$ \cite{Kleihaus:2005me})
\begin{eqnarray}
\label{scalar-pot}
U(\phi)=M^2 \phi^2-\lambda \phi^4+\nu \phi^6,
\end{eqnarray}
with $\lambda$, $\nu$ positive constants.
The total mass-energy and angular momentum of the gauged Q-balls  
are found by integrating over the entire space
the corresponding components 
 of the energy-momentum tensor,
$E=-\int d^3 x T_t^t$, 
$J= \int d^3 x T_\varphi^t$.

Before discussing the properties of the solutions, we note the existence of 
the following virial identity  
\begin{eqnarray}
\label{virial}
{\cal T}+3 {\cal U}+{\cal S}={\cal E}_M+3{\cal Q}~,
\end{eqnarray} 
 with
 \begin{eqnarray}
\nonumber
&&
{\cal T}=\int_0^{\infty}dr \int_0^{\pi}d\theta \sin^2 \theta
\bigg[
(1+\frac{r^2}{\ell^2})\phi_{,r}^2
+\frac{\phi_{,\theta}^2}{r^2}
+\frac{(n+q A_\varphi \sin\theta)^2\phi^2}{r^2\sin^2\theta}
\bigg],
\\
\nonumber
&&
{\cal U}=\int_0^{\infty}dr \int_0^{\pi}d\theta \sin^2 \theta~
U(\phi)  ,~~
{\cal Q}=\int_0^{\infty}dr \int_0^{\pi}d\theta \sin^2 \theta
\frac{(\omega-q A_t)^2\phi^2}{ 1+\frac{r^2}{\ell^2} }~, 
\end{eqnarray}

\setlength{\unitlength}{1cm}
\begin{picture}(8,6) 
\put(-0.5,0.0){\epsfig{file=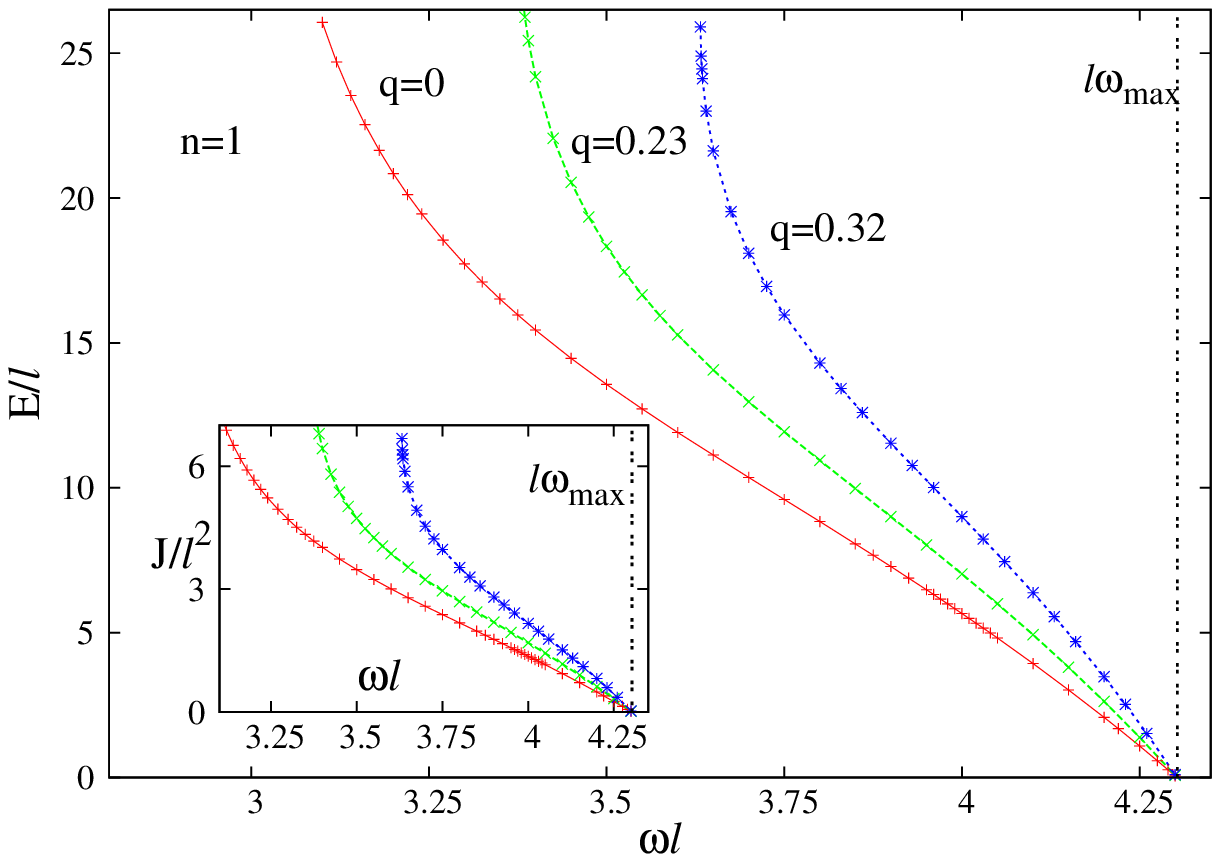,width=8cm}}
\put(8,0.0){\epsfig{file=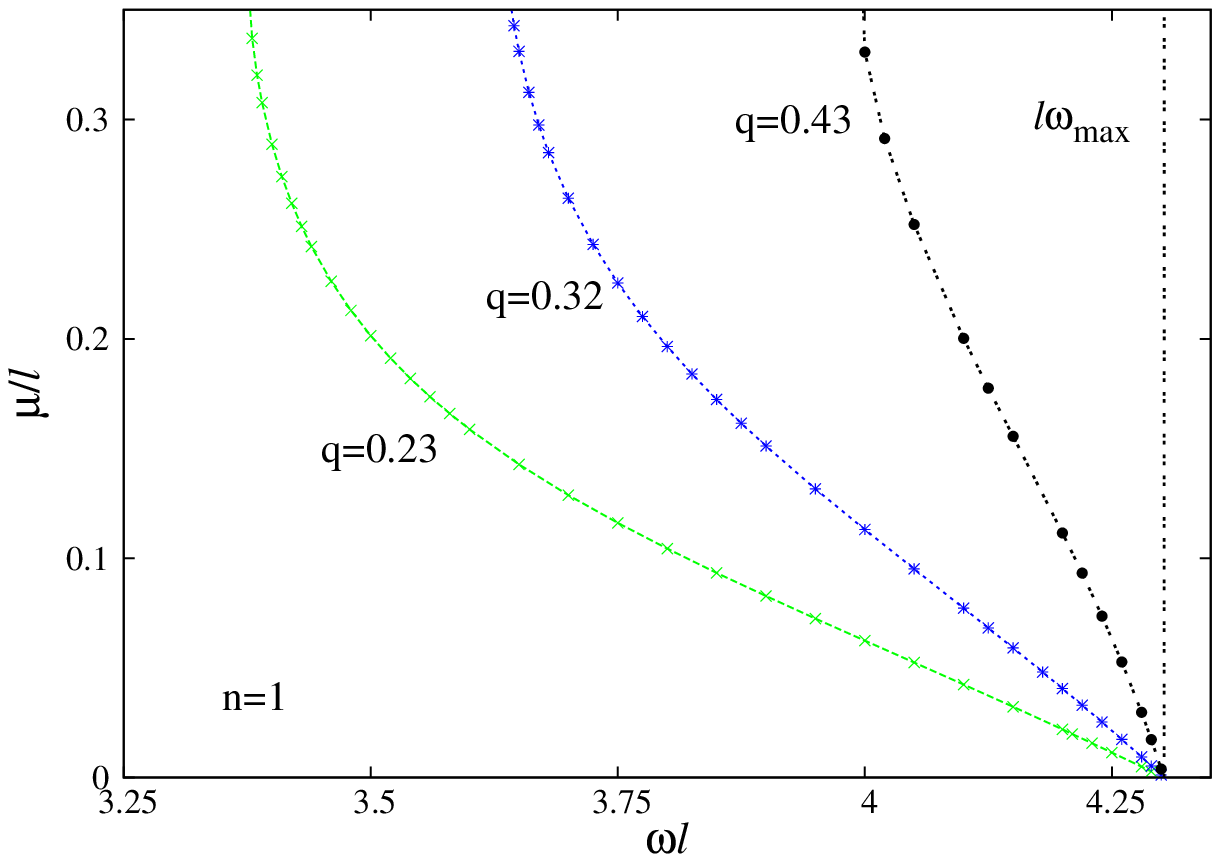,width=8cm}}
\end{picture}
\\
\\
{\small {\bf Figure 1.} 
The mass-energy $E$, the angular momentum $J$ and the magnetic dipole moment $\mu$
are shown as a function of the frequency
$\omega$ for a typical  non-gravitating Q-ball solutions with several
values of $q$.
   }
\vspace{0.1cm}
\\
\\
positive terms,
and
\begin{eqnarray}
\nonumber
{\cal E}_M=
\int_0^{\infty}dr \int_0^{\pi}d\theta \sin^2 \theta 
\bigg [
\frac{(1+\frac{r^2}{\ell^2}) A_{\varphi,r}^2}{2r^2}
+\frac{1}{2}A_{t,r}^2
+\frac{A_{t,\theta}^2}{2r^2(1+\frac{r^2}{\ell^2})}
+\frac{(A_{\varphi,\theta}+A_{\varphi})^2}{2r^4}
\bigg] \geq 0 ,
\end{eqnarray}
 a  term proportional to the total  energy stored  in the electromagnetic field.
Also
\begin{eqnarray}
\nonumber
&&
 {\cal S}=
-\frac{2\Lambda}{3}\int_0^{\infty}dr \int_0^{\pi}d\theta \sin^2 \theta
r^2
\bigg [
\phi_{,r}^2+\frac{(\omega-q A_t)^2 \phi^2}{(1+\frac{r^2}{\ell^2})^2}
+\frac{A_{\varphi,r}^2}{2r^2}
+\frac{A_{t,\theta}^2}{2(1+\frac{r^2}{\ell^2})^2r^2}
\bigg],
\end{eqnarray} 
is a (positive) term encoding the contribution of the
AdS background. 
The total mass-energy of the solutions is
$E= {\cal T}+{\cal Q}+{\cal U}+{\cal E}_M$.

One can see that the solutions with a strictly
positive potential, $U(\phi)> 0$, are supported by the harmonic time
dependence of the scalar field together with the contribution of the U(1) field.
Also, note that the identity (\ref{virial}) does not forbid the existence of static $\omega=A_t=0$
gauged configurations.
 However, similar to the case of a flat spacetime background
 \cite{Radu:2008pp}, 
 no such solutions have been found for $U(\phi)>0$.

Since the basic properties of the gauged configurations are rather similar to those of the
  Q-balls discussed in \cite{Radu:2012yx}, we shall start with a brief 
review of that case.
Supposing that $U(\phi)>0$,
the solutions exist for a limited range of frequencies, $0<\omega_{min}< \omega <\omega_{max}$,
with
\begin{eqnarray}
\label{wmax}
\omega_{max}=\frac{n+\Delta}{\ell}.
\end{eqnarray}
As $\omega\to \omega_{max}$, the Q-balls emerge  as a perturbation
around the ground state $\phi=0$, while
both the mass-energy and
Noether charge appear to diverge\footnote{Note the difference with respect to the case
of a Minkowski spacetime background.
There the  mass-energy and angular momentum
 grow without bounds as $\omega$ approaches the limits of the allowed frequency range, with $\omega_{max}=M$.
In between, there is a critical
value of the frequency, for which both $E$ and $J$ 
attain their minimal values and a double-branch structure is observed.} 
as $\omega\to \omega_{min}$.

\setlength{\unitlength}{1cm}
\begin{picture}(8,6) 
\put(-0.5,0.0){\epsfig{file=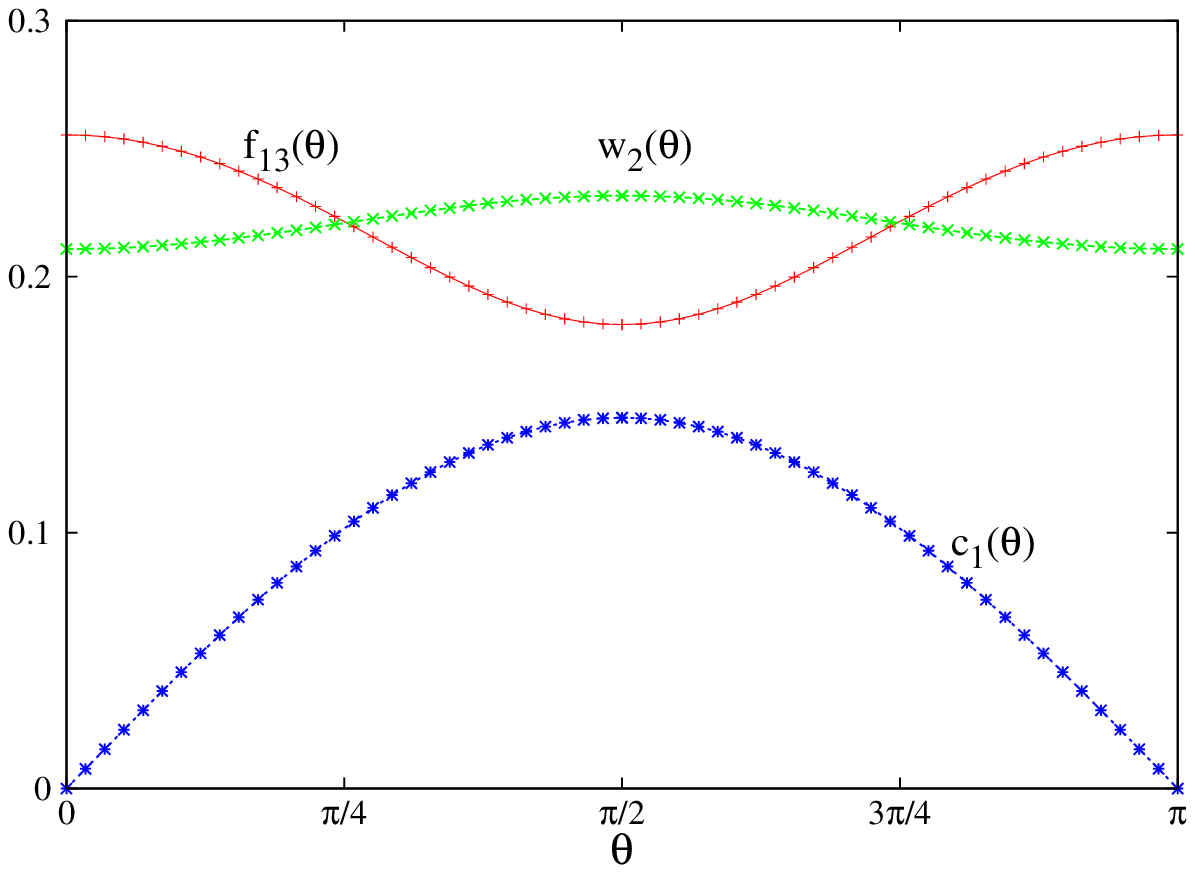,width=8cm}}
\put(8,0.0){\epsfig{file=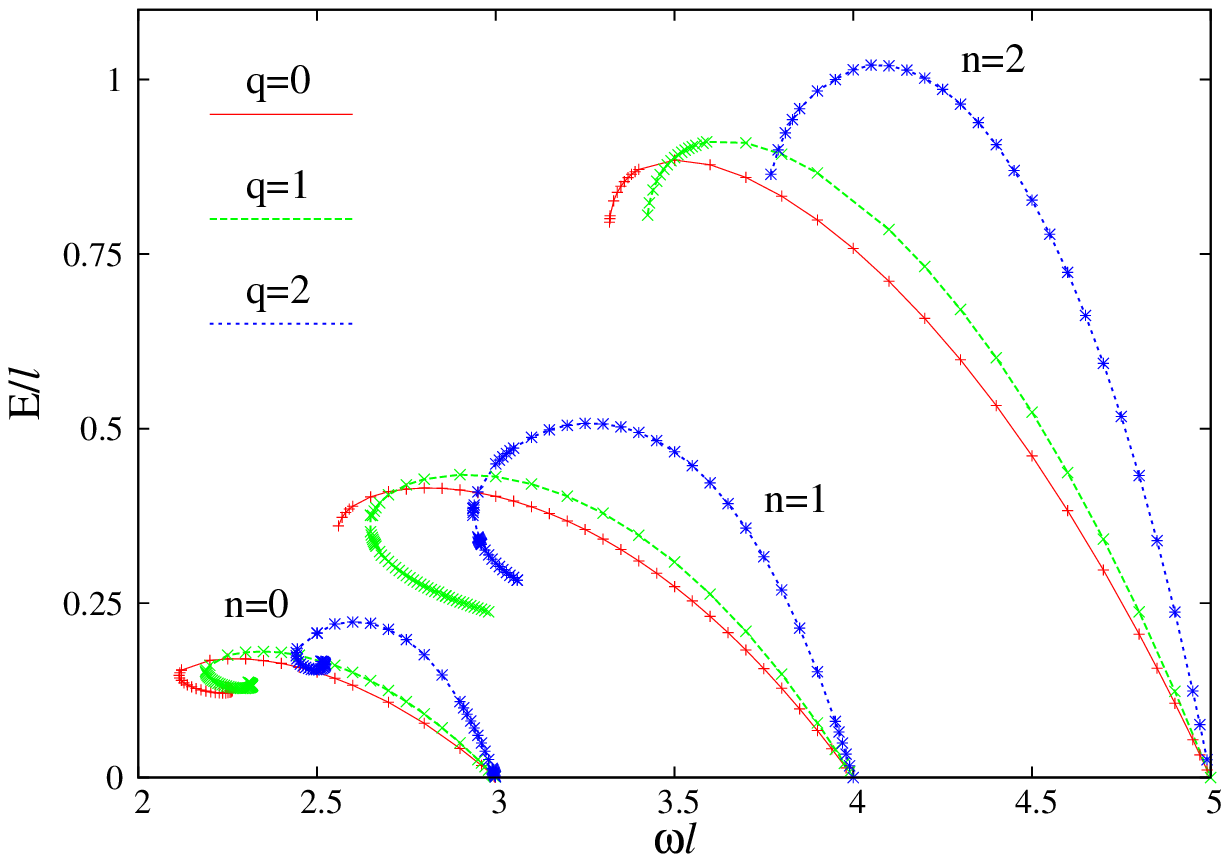,width=8cm}}
\end{picture}
\\
\\
{\small {\bf Figure 2.} 
{\it Left:} The functions $f_{13}(\theta)$, $w_2(\theta)$ and $c_1(\theta)$
which enter the large-$r$ asymptotics are shown for a typical  solution with 
$\omega \ell=3$, $q=2$, $n=1$.
{\it Right:} 
The mass-energy $E$  
is shown as a function of the frequency
$\omega $ for gauged boson star solutions with several values of the gauge coupling constant $q$.
   }
\vspace{0.1cm}
\\
 
We have found that any spinning AdS Q-ball in \cite{Radu:2012yx} 
possesses generalizations with a U(1) gauged field.
The solutions are found by slowly increasing the 
value of gauge coupling constant $q$.
For $q\neq 0$, the frequency dependence of the solutions
is similar to the Q-ball case, see Figure 1 
(the solutions there have been found for a potential of the form (\ref{scalar-pot}) with $M=1$, $\lambda=-2$, $\nu=0.2$).
The maximal value of the frequency is still given by (\ref{wmax}),
while $\omega_{min}$ decreases with $q$.

The picture found when keeping fixed all other input parameters and 
varying the value of $q$
is similar to that discussed below for the gravitating case,
with the existence of a maximal allowed value for $q$.

\subsection{Gravitating spinning gauged boson stars}

To construct gravitating solutions, we use again the ungauged spinning Q-balls
in \cite{Radu:2012yx} as seed configurations and slowly increase the value of
the gauge coupling constant $q$.
Once a solution is found with a nonzero gauge field potential,
we keep $q$ constant and couple to gravity. 

To simplify the picture, 
we restrict the discussion 
here to the case of a
scalar field with a vanishing potential\footnote{However, we have constructed also solutions with a  
scalar potential of the form (\ref{scalar-pot}).
Our results suggest that the qualitative properties of the solutions discussed here
are generic as long as $U(|\psi|)>0$.
Families of solutions with a tachyonic mass of the
scalar field, $U(|\psi|)=M^2 |\psi|^2<0 $,
have been also considered.
However, the study of such configurations requires a deviation
from the scheme proposed in this work.
For example,  the scalar field possesses in this case a more general asymptotics than (\ref{asym-matter-fields}),
which implies in general a different boundary condition in the far field region than the one in (\ref{bcinf}).  
}
$U(|\psi|)=0$.

\setlength{\unitlength}{1cm}
\begin{picture}(8,6) 
\put(-0.5,0.0){\epsfig{file=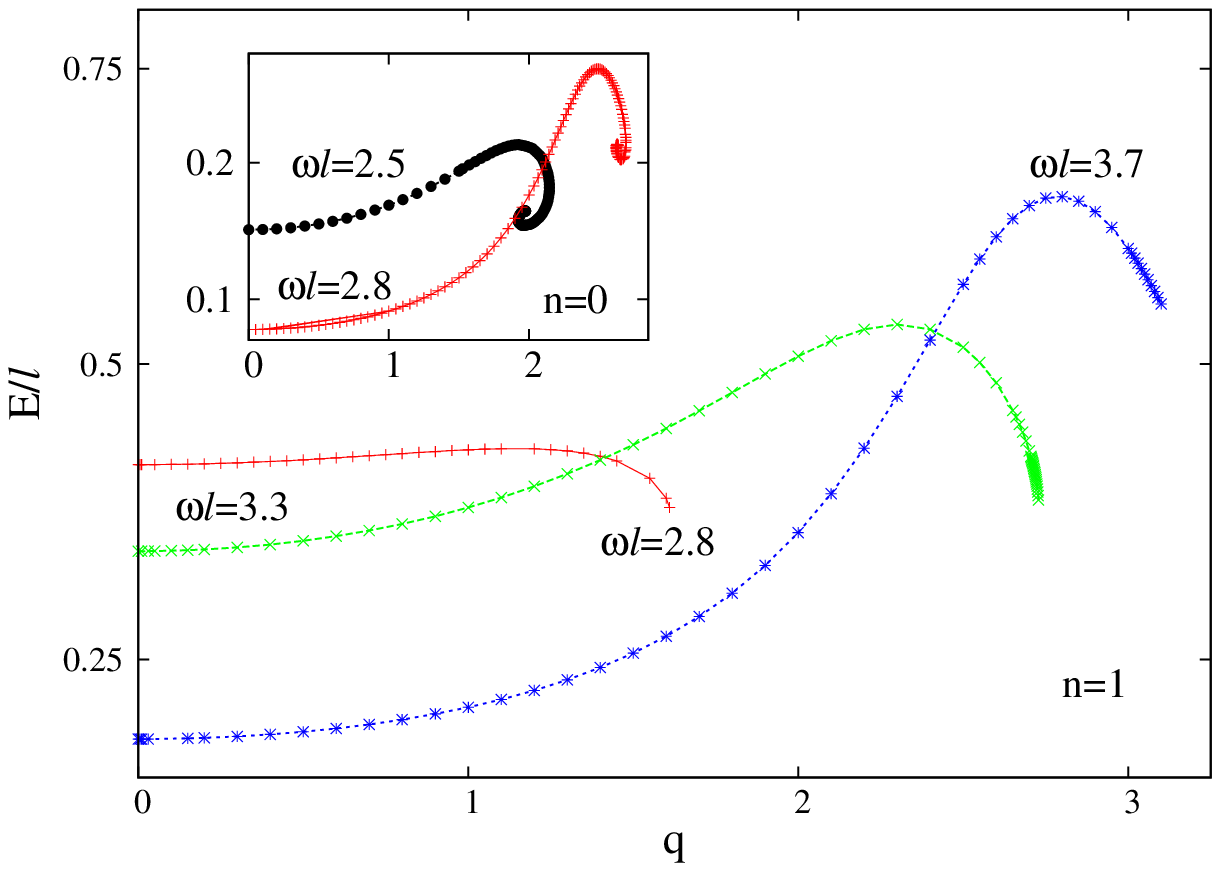,width=8cm}}
\put(8,0.0){\epsfig{file=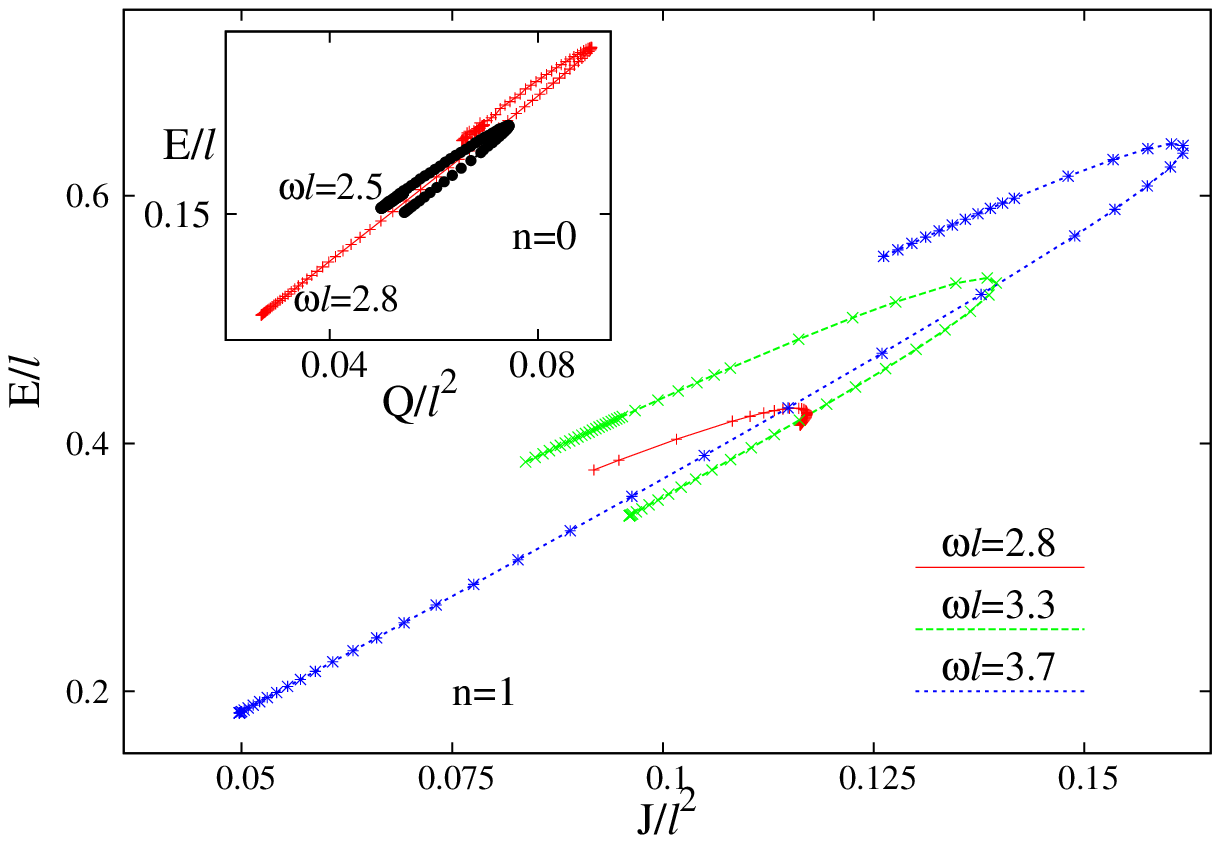,width=8cm}}
\end{picture}
\\
\\
{\small {\bf Figure 3.} 
{\it Left:} The mass-energy $E$ is shown as a function of the 
gauge coupling constant
$q$
for gauged boson star solutions with several frequencies.
{\it Right}: The $(J,E)$-diagram for the same solutions.
   }
\vspace{0.1cm}
 \\

The basic properties of these gravitating spinning gauged boson stars solutions can be summarized as follows.
First, they have $n\geq 1$,
i.e., being disconnected from the spherically 
symmetric sector, they do not possess
a slowly rotating limit. 

Second, the gauged boson stars
exist for a limited range of frequencies
$0<\omega_{min}<\omega<\omega_{max}$, emerging as a perturbation of the globally 
AdS spacetime
for a critical frequency $\omega_{max}$ as given by (\ref{wmax}).
As seen in Figure 2 (right), 
the minimal frequency increases with $q$.
In contrast to the probe limit discussed above,
the global charges stay finite as $\omega\to \omega_{min}$.
Instead,  a backbending towards
larger values of $\omega$  is observed there, see Figure 2 (right). 
One may expect that,
similar to the spherically symmetric case, this backbending would
lead to an inspiraling of the solutions towards a limiting configuration
with $\omega_c>\omega_{min}$.

Third, for given values of $\omega \ell$ and $n$,
spinning solutions exist up to a maximal
value of the gauge coupling constant only, $q=q_{max}$, see Figure 3.
The mechanism explaining this behaviour 
is similar to the spherically symmetric case.
For $q>q_{max}$ the charge repulsion  becomes bigger than
  the gravitational attraction and the localized solutions cease to exist
  (this feature has been noticed already in the initial study \cite{Jetzer:1989av}
  of the spherically symmetric, asymptotically flat charged boson stars).
As seen in Figure 3, the maximal value of $q$
increases with frequency.
Also, all global charges stay finite as $q\to q_{max}$.
Unfortunately, the numerical
accuracy does not allow to clarify the limiting behavior\footnote{Based on the results for $n=0$
(see the inset in Figure 3 (left)),
we expect a complicated behaviour,
with the occurrance of secondary branches and
 an inspiraling in $q$
 towards a limiting configuration
with $q_c<q_{max}$.
}
 at the
maximal value of $q$. We notice only that, as $q\to q_{max}$, the metric
function $F_0$ takes very small values at $r = 0$,

\setlength{\unitlength}{1cm}
\begin{picture}(8,6) 
\put(-0.5,0.0){\epsfig{file=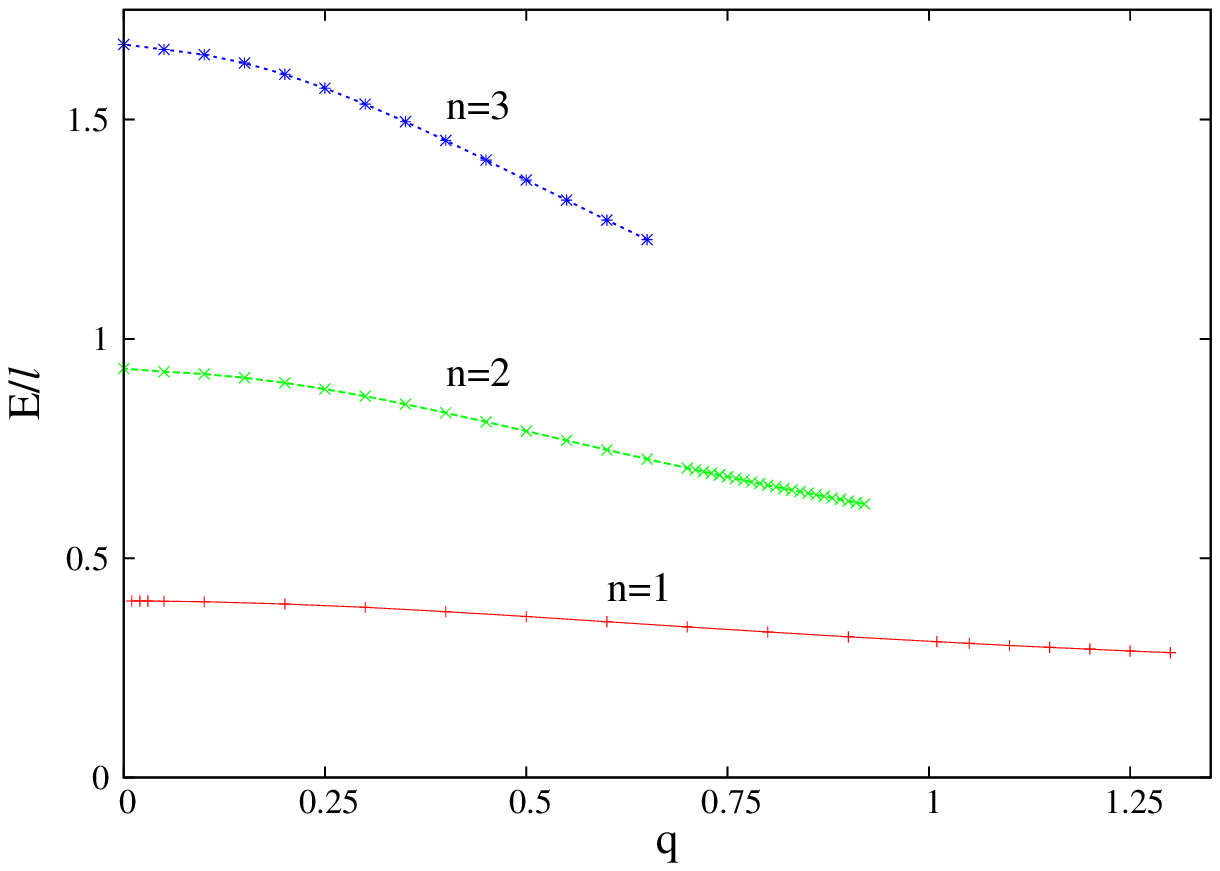,width=8cm}}
\put(8,0.0){\epsfig{file=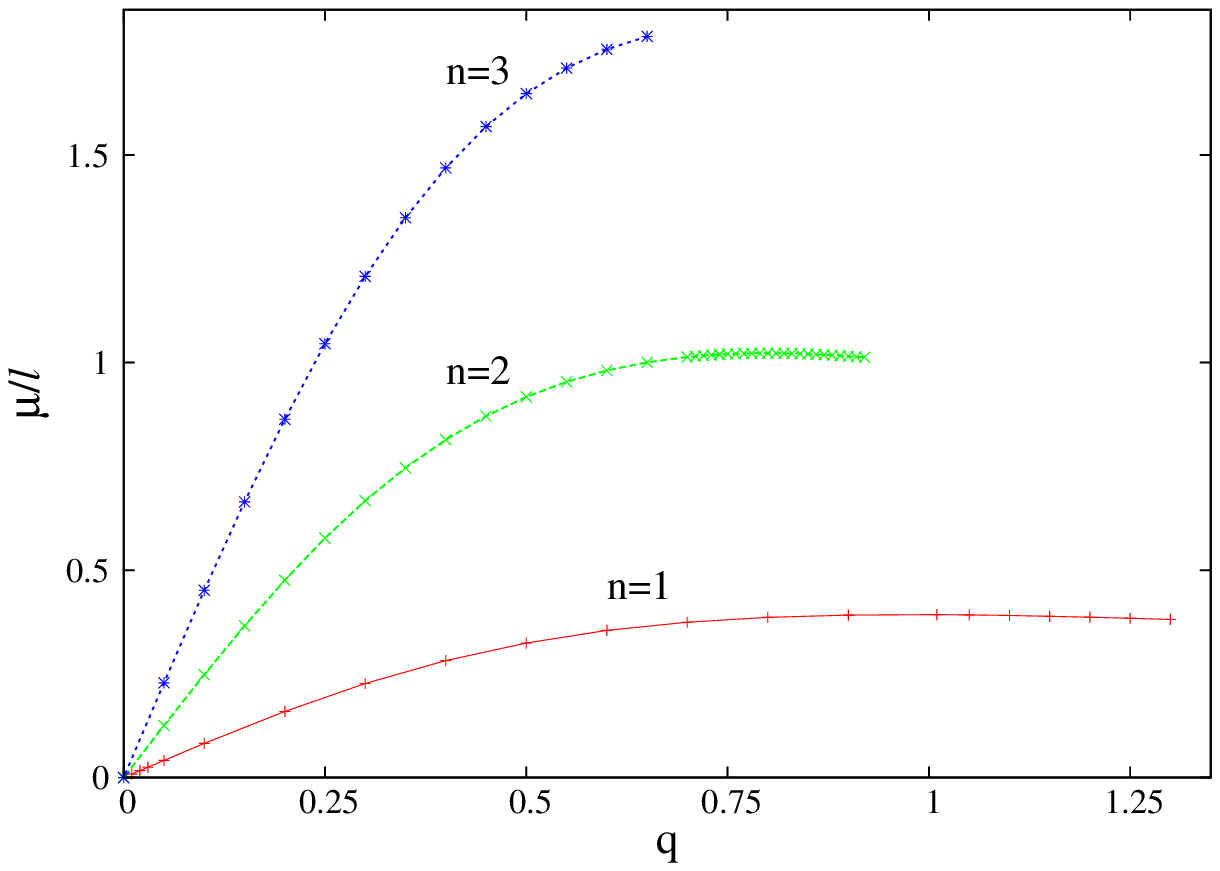,width=8cm}}
\end{picture}
\\
\\
{\small {\bf Figure 4.} 
The mass-energy $E$ and the magnetic moment $\mu$ are shown as functions of the 
gauge coupling constant
$q$ for static axially symmetric solutions with several winding numbers.
   }
\vspace{0.1cm}
\\
\\
 while the other functions
remain finite and nonzero (although $F_1$ and $F_2$ take large values
at the origin).

Fourth, the shape of the metric functions and of the scalar field $\phi$
is rather similar to the ungauged case.
Concerning the gauge field, the electric potential $A_t$
does not possess a 
 strong angular dependence; 
however the magnetic potential $A_\varphi$
exhibits a complicated angular dependence.
The energy density of all solutions
has a strong peak in the equatorial plane
and decreases monotonically along the symmetry axis,
such that the typical energy
density isosurfaces have a toroidal shape. 
As seen in Figure 2 (left), the functions 
$f_{12}$, 
$w_2$,
and
$c_1$
which enter the large-$r$ asymptotics,
have a strong $\theta$-dependence.

\subsection{The static limit: gauged scalar solitons with magnetic dipole moment}

An interesting property of a set of configurations considered 
in this work is the possible existence of a nontrivial
 limit, describing static gauged solutions
 with a magnetic dipole moment and no electric field\footnote{To our knowledge,
all known particle-like solutions with this property exist 
in models with non-Abelian fields,
the electroweak sphaleron with a nonvanishing mixing angle \cite{Kleihaus:1991ks}
being perhaps the best known
example.} 
(thus $A_t=W=0$ in this case).

These solutions are supported by 
a field potential which is not strictly positive definite
and exist already in the probe limit, 
see the virial relation (\ref{virial}).
We have studied such solutions for several choices of the parameters in 
the general potential (\ref{scalar-pot})
(the solutions in Figure 4 have 
$U(|\psi|)=-\lambda |\psi|^4$,
with $\lambda=2.2$).

The limit $q=0$ corresponds to the static axially symmetric 
scalar solitons whose existence has been reported in \cite{Radu:2012yx}.
Again, the gauged solutions emerge from these configurations
by slowly increasing the value of the gauge coupling constant. 
As seen in Figure 4, in contrast to the spinning case, the mass
of the static solutions decreases with frequency.
 In the probe limit, this can be understood as follows. 
For small values of $q$, one can write the expansion
 $
  \phi=\phi^{(0)}+q^2 \phi^{(2)}+O(q^4),~~ A_{\varphi}=q A_{\varphi}^{(1)}+ O(q^3),
 $
 where $\phi^{(0)}$ is the ungauged configuration.
After inserting this into the expression of the total mass-energy and using the equations of motion (\ref{scalar-eqs}),
  one finds
 $E=E^{(0)}-q^2 E^{(2)}$, 
 where
 $E^{(2)}=\frac{1}{4}\int d^3 x F_{\mu\nu}^{(1)\mu \nu}$ and $E^{(0)}$ the total mass-energy of the ungauged spinning boson star.
 
Although the above physical argument for the existence of a maximal 
value of $q$ does not apply for these static solutions
(since the electric repulsion is absent), 
the numerical results suggest that 
a maximal value of $q$ does nevertheless exist, see Figure 4.
As $q\to q_{max}$, the critical solutions\footnote{Note, however, that finding an accurate value  for  $q_{max}$
has proven a difficult task 
(at least for the input
parameters we have considered), 
since the numerical accuracy
decreases close to that point, yielding relatively large violations for the constraint equations.} exhibit
the same behaviour as the one noticed in the spinning case. 
In particular, the relevant quantities remain finite close to that point,
while $F_0(0,0)$ and $1/F_1(0,0)$ take small values.
Also, the maximal value of $q$ decreases again as the value of the winding number $n$ increases.

\section{Further remarks}

The main purpose of this work was to propose a numerical scheme for the study of 
spinning gauged  boson stars in a four-dimensional AdS spacetime,
together with a  discussion of the basic properties of
these configurations.
In our approach, the gauged spinning boson stars emerge smoothly from the corresponding
 solutions with a vanishing electromagnetic field
 by slowly increasing the value of the gauged coupling constant.
Then, as expected, their basic properties
are rather similar to those of the ungauged configurations.
However, the gauged coupling constant cannot be arbitrarly large.

It would be desirable to study these solutions also from a
holographic dual point of view.
On general grounds,
one expects the  $d=4$ spinning gauged boson stars
 to describe zero temperature states of a
 conformal field theory (CFT) defined in 
 a fixed background
with $ds^2=\gamma_{ab}dx^a dx^b=-dt^2+\ell^2(d\theta^2+\sin^2 \theta d\varphi^2)$
($i.e.$ a $2+1$ Einstein universe).
As usual, the asymptotic behavior of the bulk solutions determines certain
properties of the dual field theory.
%
For example, the asymptotics of the temporal component of the gauge potential 
(in a gauge without a time dependence of the scalar field)
gives the charge density and the chemical potential of the CFT.
Likewise, the field $\psi$ is dual to an operator ${\cal O}$
in the CFT, with a scaling dimension $\Delta$.
One can also compute the expectation value of the dual CFT stress-tensor $<\tau_{ab}>$ 
by
using the relation \cite{Myers:1999qn}
$
\sqrt{-\gamma}\gamma^{ab}<\tau_{bc}>=
\lim_{r \rightarrow \infty} \sqrt{-h} h^{ab}{\rm T}_{bc}
$
(with ${\rm T}_{ab}$ the boundary stress tensor).
This results in the interesting expression
(with 
$x^1=\theta$, 
$x^2=\varphi$,
$x^3=t$)
\begin{eqnarray}
\nonumber
8\pi G \ell^4 <\tau^{a}_b>= 
 A_1(\theta) \pmatrix{1&0&0\cr 0&1&0\cr 0&0&-2\cr} +
 A_2(\theta) \pmatrix{1&0&0\cr 0&3&0\cr 0&0&-4\cr} +
 A_3(\theta) \pmatrix{0&0&0\cr 0&0&\ell^2\sin^2\theta\cr 0&-1&0\cr},
\end{eqnarray}
(where 
$A_1(\theta)=2f_{13}-\tan \theta f_{13}'$,
$A_2(\theta)=\tan \theta f_{13}'$,
$A_3(\theta)=\frac{3}{2}w_2$),
which is finite, covariantly conserved and manifestly traceless.

The inclusion of finite temperature effects in this model would require 
to construct the corresponding black hole solutions.
In this context, the numerical study 
in this work can  be viewed as a necessary step before approaching
the more complex case of a spinning black hole of the 
gravitating Abelian-Higgs model in a globally AdS background\footnote{Spinning solitons and 
black hole solutions of the 
gravitating Abelian-Higgs model have been constructed in  \cite{Brihaye:2011fj}
for a special metric ansatz with two equal angular momenta and an AdS$_5$ background.}.
Different from the ungauged case, the model (\ref{action})
possesses black hole solutions with  scalar hair
already in the spherically symmetric limit, see $e.g.$ \cite{Gentle:2011kv}.
The numerical results in Ref. \cite{Sonner:2009fk}
imply that these static black holes with scalar hair possess
also spinning generalizations\footnote{This follows from the explicit construction 
in \cite{Sonner:2009fk} of a marginal
mode of the Kerr-Newman-AdS black hole. 
This implies the existence of a new branch of charged stationary solutions
with a nonzero scalar condensate outside the horizon.
However, no explicit construction of these solutions 
has been reported so far in the literature.}.
Then, on general grounds, we expect
that when putting  together  the solitons in this work
with the (yet to be found) hairy black holes,
 the resulting picture  
will be similar to that revealed 
in \cite{Dias:2011at}
for a version of the model without 
gauge fields.
That is, we conjecture that when the horizon size shrinks to zero, the hairy black hole solutions of the gravitating Abelian-Higgs model
will reduce to the
rotating gauged boson stars discussed above.
Conversely, one can put a small rotating black hole inside any spinning, horizonless solution 
in this work.
The angular velocity of the horizon of a rotating black hole is  
$ \Omega_H=(\omega-q \Phi_H)/n$,
 with $\Phi_H$  the electrostatic potential on the horizon.
We hope to return elsewehere with a systematic study of these aspects.

Finally, let us remark that the  rotating gauged boson stars
studied in this paper can be considered as
simple prototypes of more complicated spinning configurations,
possibly with non-Abelian gauge fields.
Moreover, they may provide as well a fertile ground for further study of charged rotating configurations
in gauged supergravity models, with a more complicated action than (\ref{action}).
Thus the study in this work gives an idea of how a similar
procedure would work in those cases.
Also, we expect that the  numerical scheme proposed here 
could also be applied to the study of higher dimensional solutions of the
gravitating Abelian Higgs model with rotation in a single plane.

\vspace*{0.5cm}

\noindent{\textbf{~~~Acknowledgements.--~}}  
We gratefully acknowledge support by the DFG,
in particular, also within the DFG Research
Training Group 1620 ``Models of Gravity''. 
 \begin{small}
 
 \end{small}
 
 \end{document}